\documentclass[twocolumn,preprint2]{aastex63}
\usepackage{graphicx,times,epsf}    
\usepackage{soul,xcolor}
\usepackage{longtable}
\usepackage{natbib}

\setstcolor{red}                                  
\shortauthors{Yang et al.}
\begin{document}
\title{LTD064402+245919: A Subgiant with a 1-3 M$_{\odot}$ Undetected Companion Identified from LAMOST-TD Data} 
\email{bozhang@bnu.edu.cn (Bo Zhang)}
\email{shansusu@nao.cas.cn (Su-Su Shan)}

\author[0000-0002-6039-8212]{Fan Yang}
\affil{Department of Astronomy, Beijing Normal University, Beijing 100875, People's Republic of China\\}
\affil{National Astronomical Observatories, Chinese Academy of Sciences, 20A
Datun Road, Chaoyang District, Beijing 100101, China\\}
\affil{School of Astronomy and Space Science, University of Chinese Academy of Sciences,
Beijing 100049, China\\}
\author[0000-0002-6434-7201]{Bo Zhang}
\affil{Department of Astronomy, Beijing Normal University, Beijing 100875,
People's Republic of China\\}
\author[0000-0002-8559-0067]{Richard J. Long}
\affil{Department of Astronomy, Tsinghua University, Beijing 100084, China\\}
\affil{Jodrell Bank Centre for Astrophysics, Department of Physics and Astronomy, The University of Manchester, Oxford Road, Manchester M13 9PL, UK\\}
\author[0000-0002-1310-4664]{You-Jun Lu}
\affil{National Astronomical Observatories, Chinese Academy of Sciences, 20A
Datun Road, Chaoyang District, Beijing 100101, China\\}
\affil{School of Astronomy and Space Science, University of Chinese Academy of Sciences,
Beijing 100049, China\\}
\author[0000-0002-5744-2016]{Su-Su Shan}
\affil{National Astronomical Observatories, Chinese Academy of Sciences, 20A
Datun Road, Chaoyang District, Beijing 100101, China\\}
\affil{School of Astronomy and Space Science, University of Chinese Academy of Sciences,
Beijing 100049, China\\}
\author {Xing Wei}
\affil{Department of Astronomy, Beijing Normal University, Beijing 100875,
People's Republic of China\\}
\author {Jian-Ning Fu}
\affil{Department of Astronomy, Beijing Normal University, Beijing 100875,
People's Republic of China\\}
\author {Xian-Fei Zhang}
\affil{Department of Astronomy, Beijing Normal University, Beijing 100875,
People's Republic of China\\}
\author{Zhi-Chao Zhao}
\affil{Department of Astronomy, Beijing Normal University, Beijing 100875,
People's Republic of China\\}
\author{Yu Bai}
\affil{National Astronomical Observatories, Chinese Academy of Sciences, 20A
Datun Road, Chaoyang District, Beijing 100101, China\\}
\author[0000-0002-5839-6744]{Tuan Yi}
\affil{Department of Astronomy, Xiamen University, Xiamen, Fujian 361005, P. R. China\\}
\author[0000-0002-5630-7859]{Ling-Lin Zheng}
\affil{Department of Astronomy, Xiamen University, Xiamen, Fujian 361005, P. R. China\\}
\author{Ze-Ming Zhou}
\affil{Department of Astronomy, Beijing Normal University, Beijing 100875,
People's Republic of China\\}
\author[0000-0002-2874-2706]{ji-feng liu}
\affil{National Astronomical Observatories, Chinese Academy of Sciences, 20A Datun Road, Chaoyang District, Beijing 100101, China\\}
\affil{School of Astronomy and Space Science, University of Chinese Academy of Sciences,
Beijing 100049, China\\}
\affil{WHU-NAOC Joint Center for Astronomy, Wuhan University, Wuhan, China\\}

\begin{abstract}
Single-line spectroscopic binaries recently contribute to the stellar-mass black hole discovery, independently of the X-ray transient method. We report the identification of a single-line binary system LTD064402+245919, with an orbital period of 14.50 days. The observed component is a subgiant with a mass of 2.77$\pm$0.68M$_{\odot}$, radius 15.5$\pm$2.5R$_{\odot}$, effective temperature $T_{\rm eff}$ 4500$\pm$200K, and surface gravity log\emph{g} 2.5$\pm$0.25dex. The discovery makes use of the LAMOST time-domain (LAMOST-TD) and ZTF survey. Our general-purpose software pipeline applies the Lomb-Scargle periodogram to determine the orbital period and uses machine-learning to classify the variable type from the folded light curves. We apply a combined model to estimate the orbital parameters from both the light and radial velocity curves, taking constraints on the primary star mass, mass function, and detection limit of secondary luminosity into consideration. We obtain a radial velocity semi-amplitude of 44.6$\pm$1.5 km s$^{-1}$, mass ratio of 0.73$\pm$0.07, and an undetected component mass of 2.02$\pm$0.49M$_{\odot}$ when the type of the undetected component is not set.  We conclude that the inclination is not well constrained, and that the secondary mass is larger than 1M$_{\odot}$ when the undetected component is modelled as a compact object. According to our investigations using an MCMC simulation, increasing the spectra SNR by a factor of 3 would enable the secondary light to be distinguished (if present). The algorithm and software in this work are able to serve as general-purpose tools for the identification of compact objects quiescent in X-rays.
\end{abstract}

\section{Introduction}
Optical spectroscopic binaries quiescent in X-ray have recently attracted growing attention due to the potential for stellar-mass black hole and neutron star discoveries.
The traditional and commonly used identification of stellar-mass black holes (BHs) and neutron stars (NSs) is via transient X-ray emission, due to significant gas accretion \citep{liuxrayBH, Kreidberg2012}. The number of confirmed X-ray identified BHs is $\sim$ 20, with more recent searches utilizing radial velocity (RV) modeling \citep{lb1liu, liulb12020, Thompson2019}. Radial velocity identified systems are in a quiescent phase with very faint (even undetectable) X-ray emissions. The likely targets are expected to have computable binary parameters and have a large undetected mass. Stellar evolution models predict the existence of $\sim$ 10$^{9}$ BHs in the Milk Way \citep{Brown1994, Timmes1996}, though only about 100 of them (including candidates) have been found \citep{Corral-Santana2016}. X-ray quiescent BHs are expected to significantly enlarge the BH sample \citep{Masuda2019,Lu2020,Clavel2020}.

A binary system with an undetected massive object is potentially effective for searching for BHs and NSs.
\citet{gu2019} proposed a method for detecting compact objects with giant stars with the Large Sky Area Multi-Object fiber Spectroscopic Telescope \citep[LAMOST;][]{LAMOST}. Attempts are ongoing to search for stellar-mass black holes and neutron stars through spectroscopic surveys \citep{Lu2020, Zheng2019,yituan2019}.

In 2016, a time-domain spectroscopic survey was initiated with LAMOST for the specific purpose of finding compact objects \citep[LAMOST-TD;][]{liulb12020}. This survey has now monitored more than 10000 targets with a cadence of one week.

In this paper, we report a single-line binary LTD064402+245919, consisting of a subgiant star and an undetected object of 2.02$\pm$0.49M$_{\odot}$, using the data from LAMOST-TD survey \citep[][Wang et al. in pre.]{liulb12020} and the Zwicky Transient Facility (ZTF) survey \citep{ZTF}. The paper is organized as follows. In section 2, we briefly describe the surveys and our identification procedure. In section 3, we present our analysis of the data and the binary properties obtained. We discuss and summarize our work in sections 4 and 5.

\section{Identification of LTD064402+245919 from LAMOST-TD}

\subsection{Data used}

\begin{figure}[htb]
  \centering
 \includegraphics[width=4in,height=4in]{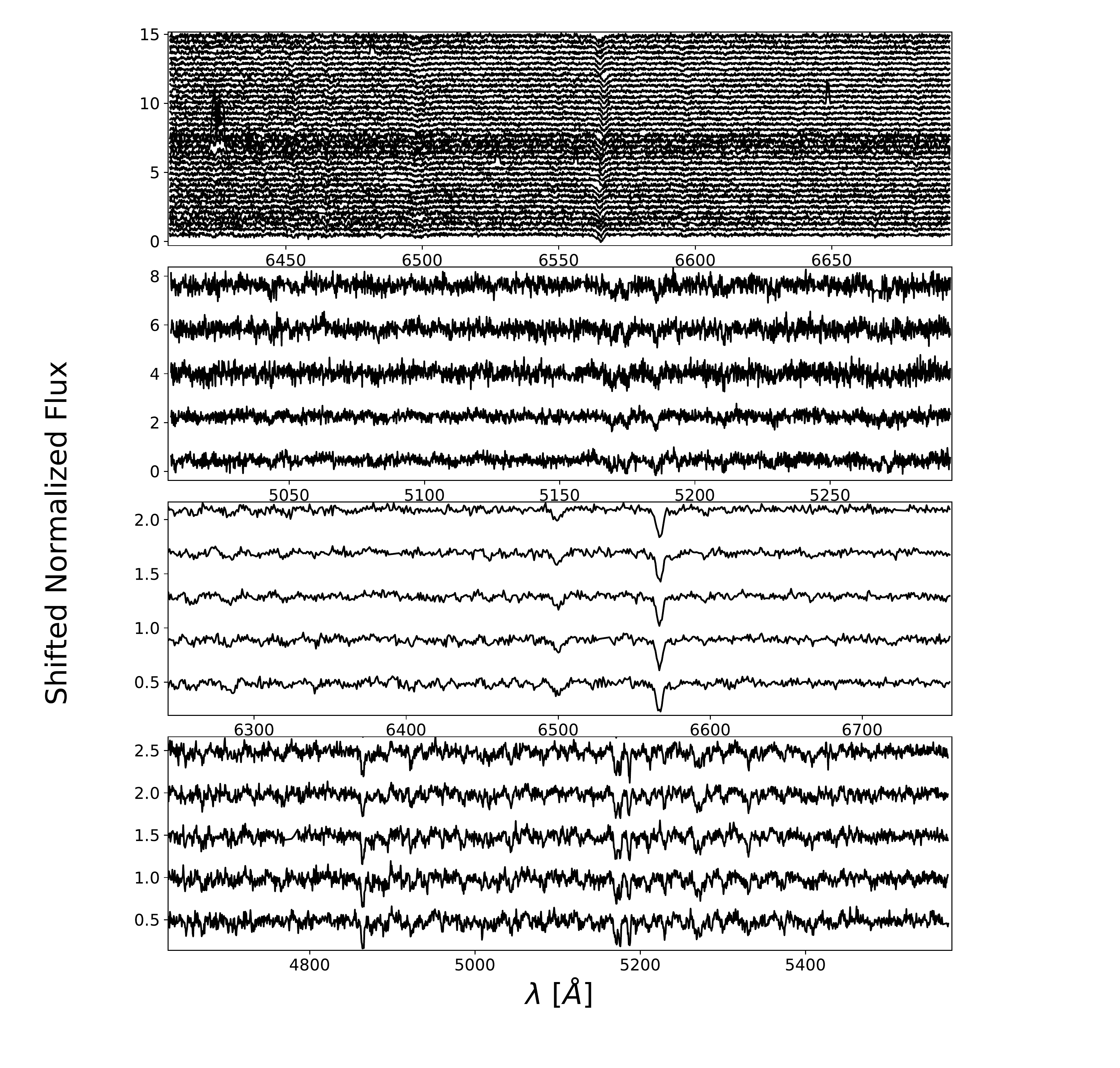}
  \caption{The vacuum-wavelength spectra of LTD064402+245919. The panels from top to bottom indicate MR, MB, LR, LB. The top panel presents all the MR spectra we used, while the rest of the panels only present the first 5 observed spectra for clarity. The H$\alpha$ lines (6562.8 \AA) in the top panel reveal obvious shifts among different observing times. The spectra median signal-to-noise ratio (SNR) from top to bottom panels are 12, 6, 28, 13.}
\label{image:spectra}
\end{figure}

LTD064402+245919 was discovered using LAMOST-TD and ZTF survey data. The LAMOST-TD data obtains 107 usable radial velocity measurements, with $\sim$ 6 observations per observing night, as shown in Fig. \ref{image:spectra}. ZTF data provides 101 r band photometric measurements and 94 g band measurements, with 1-2 observations per night, as shown in Fig. \ref{image:lc}.

LAMOST-TD monitors stars in the Kepler K2 field, with 3-9 exposures per observing night (Wang et al. in pre.). The time-domain spectra are mainly used for finding stellar-mass black holes through radial velocity measurements \citep{lb1liu,liulb12020}. Possible candidates are examined with further analysis and follow-up observations.

Possessing 4000 fibers, LAMOST takes thousands of spectra simultaneously, in both medium resolution and low resolution modes \citep{LAMOST, Xiang2015}. The medium resolution observations have R $\sim$ 7500 and a limiting magnitude of about G=15 mag \citep{zong2020}. The low resolution observation mode has a limiting magnitude of 17.8 (r band) with R $\sim$ 1800 \citep{LAMOST}. The wavelength range of the low resolution spectra is from 370 nm to 900 nm (blue arm as LB; red arm as LR). The blue arm of the medium resolution spectra (MB) is from 495 nm to 535 nm, and the red arm (MR) from 630 nm to 680 nm. 
The exposure time is 600s for low resolution spectra and 1200s for medium resolution spectra.

ZTF is a time-domain photometric survey performed on the 48 inch Samuel Oschin telescope at Palomar Mountain \citep{ZTF}. ZTF was commissioned in 2018 and covers the whole sky with a cadence of three days to a depth of 20.5 mag.

\subsection{LTD064402+245919 identification}

Variable stars among the LAMOST-TD input sources are classified by analyzing their photometric light curves (LCs). The LCs are taken from ZTF \citep{ZTF}, the All-Sky Automated Survey for Supernovae \citep[ASAS-SN;][]{ASASVaria}, and the Catalina Sky Survey \citep[CSS;][]{CSS2009}. Full details of the identification of variable types are present in the LAMOST-TD overview (Wang et al. in pre.). Here, we just summarize them briefly.

The identification of variable stars is built into our software pipeline and involves two steps. The first is to determine their periods by applying the Lomb-Scargle method \citep{ VanderPlas2015}. The second is to analyze their LCs, removing any artifacts and classifying variable stars into, for example, eclipsing binary (EB), RR Lyrae (RRL), $\delta$ Scuti (DSCT).

\begin{figure}[htb]
  \centering
 \includegraphics[width=3.5in]{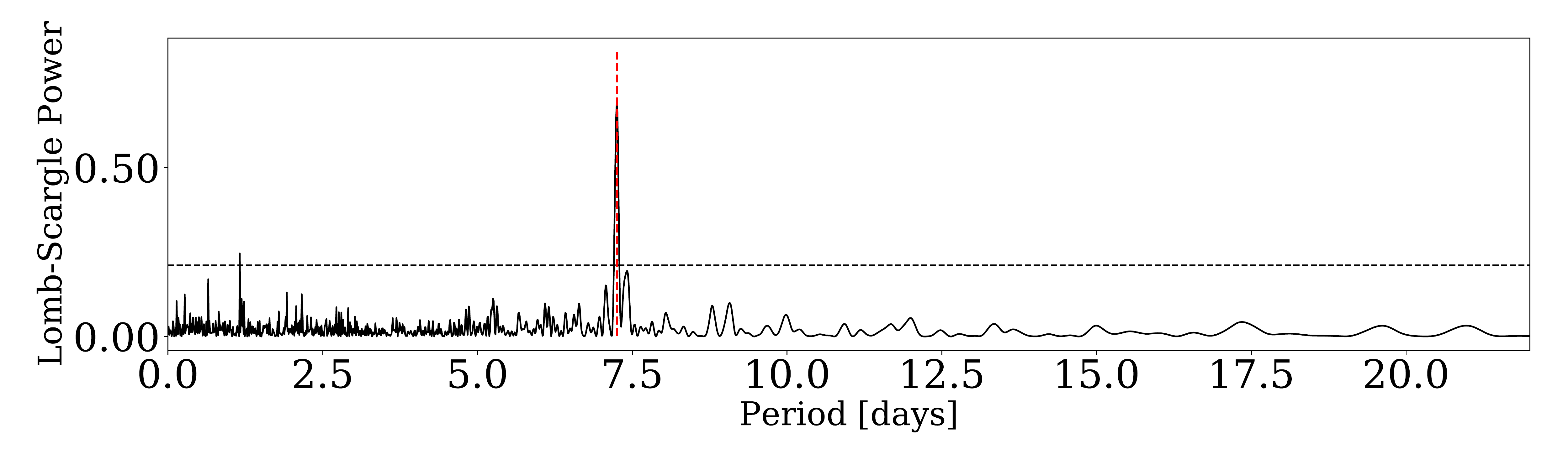}
 \includegraphics[width=3.5in]{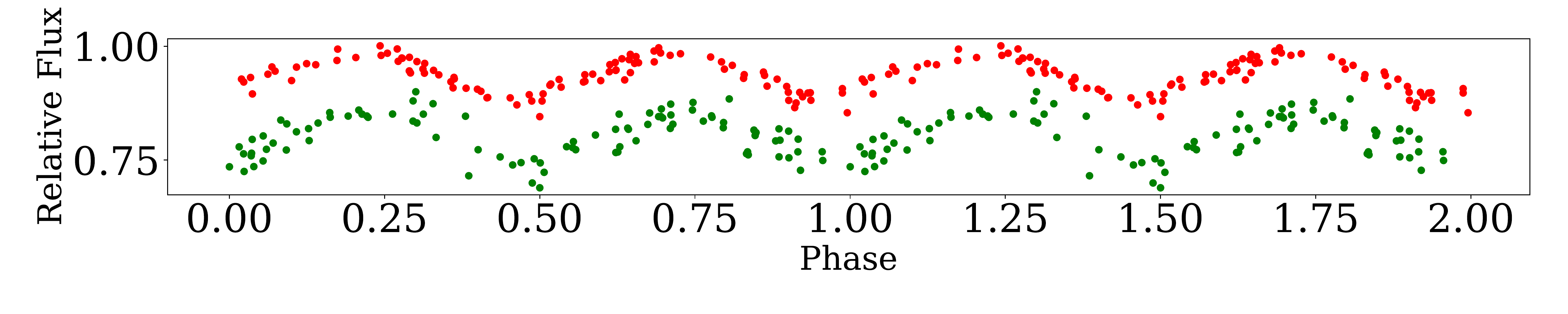}
  \caption{The Lomb-Scargle periodogram (upper) and folded light curves (lower). The red vertical line indicates the half orbital period in the periodogram. The light curves (r band in red, g band in green) are folded based on the orbital period obtained from the periodogram.}
  \label{image:lc}
\end{figure}

The significant period is identified as the period at the peak intensity in the Lomb-Scargle periodogram \citep{yangbinary}. This period is obtained from the two-step grid peak searching method \citep{VanderPlas2015}. 
The orbital period is twice the period peak in the periodogram \citep{yangbinary}. The period is regarded as significant only when the power density is greater than the false alarm threshold (as shown in Fig \ref{image:lc}, top panel). The false alarm threshold is defined as greater than the median value of the period power at 3 times the period power standard deviation. We regard the derived period of 1 day and its harmonic periods as artifacts due to the inevitable alternation between night and day for the ground-based telescope.

Light curves are folded at their significant periods, and are then compared with template light curves. The variable star templates are taken from previously created catalogs \citep[e.g.][]{GCVS2017,kim2014}. The template characteristics are included as identification parameters. These parameters are used for classification both through a machine learning method (random forest) and visual inspection \citep{long2012,kim2016,asas5,yangbinary}. The parameters include, e.g., light curve period, skewness, median, standard deviation of the magnitude distribution, the magnitude amplitude, the Fourier transform components a$_{2}$ and a$_{4}$, 10$\%$ and 90$\%$ percentile of slopes of the folded light curve. The Fourier transform of the light curve is sensitive to the classification, especially the coefficients a$_{2}$ and a$_{4}$ \citep{Paczynski2006, Ngeow2021}.

LTD064402+245919 is identified by the pipeline as a source of interest, and is categorized as a contact binary from ZTF light curves. An ellipsoidal LC is revealed with an orbital period of 14.5 days, using the archive photometric light curve from ZTF \citep{ZTF}. The Lomb-Scargle periodogram predicts a half orbital period of 7.26 days, as shown in Fig. \ref{image:lc}, where we also present the folded ZTF light curves in r band (102 measurements) and g band (98 measurements).

The raw observed data are reduced using the standard LAMOST pipeline \citep{LAMOST}. The radial velocities (RVs) are derived by cross-correlating with synthetic spectral templates \citep{Zhang2021}. The effective temperature ($T_{\rm eff}$), the surface gravity (log\emph{g}), and metallicity ([Fe/H]) are determined from the LAMOST pipeline and SLAM \citep[a machine learning method, ][]{ 2020RAA....20...51Z, Zhang2020}. The uncertainties of the parameters are taken from the statistical errors of the LAMOST samples\footnote{\url{http://www.lamost.org/dr8/v1.0/doc/release-note}} with the parameter distributions being analyzed in different SNR bins. Also, we calculate the standard deviations (stds) of LTD064402+245919 parameters among different observations. Taking log\emph{g} as an example, the uncertainties obtained from the medium resolution sources are 0.08dex, and the uncertainties obtained from the low resolution sources are 0.25dex, while the std of LTD064402+245919 log\emph{g} among different observations is 0.2dex. We take the log\emph{g} error as the largest uncertainty value from the three different statistics which is 0.25dex.

All the spectra we use are from the survey data with clipping of abnormal spectra. We omit spectra and results when the radial velocity difference obtained from the red camera and the blue camera is larger than 50 km s$^{-1}$. Also, if the modeling of the line profile fails, we reject the RV output. This clipping reduces the number of RV measurements to 107 from 120.

The LAMOST-TD radial velocity curve is folded with the period obtained from the ZTF LC, as shown in Fig. \ref{image:result}. The radial velocity curve shows a cycle frequency a factor of two smaller than the LC frequency. This helps to rule out possible false LC classifications, e.g., contaminated by stellar spots or symmetrically shaped pulsating stars.

\section{Properties of LTD064402+245919}

LTD064402+245919 has an RA and Dec of 101.0089 and 24.9887 deg. Gaia reports a G band magnitude of 14.50, with a parallax of 0.15$\pm$0.02mas \citep{GaiaDr2}.
LAMOST pipeline and SLAM consistently report that the detected companion has $T_\mathrm{eff}$, $\log{g}$, and [Fe/H] of 4500$\pm$200\,K, 2.50$\pm$0.25\,dex, and $-0.54\pm$0.18\,dex, respectively. The stellar parameters derived from the spectra strongly imply a subgiant star. Hereafter we term the observed object as the primary (observed) component and the other as the secondary (undetected) component.

The mass of the subgiant ($M_{1}$) is 2.77$\pm$0.68M$_{\odot}$, obtained from the observed luminosity, parallax, $T_{\rm eff}$, log $g$. The mass is constrained by $L \propto R^{2}T_\mathrm{eff}^{4}$, and $M$ $\propto$ $gR^{2}$, where $L$ is Luminosity of the star, $M$ its mass, and $R$ its radius. The extinction is obtained from the grid relation between direction, distance, and extinction \citep{GreenExtinction, shan2018}. \citet{GaiaDr2} report a G band extinction of 0.34 and we use this value in our calculations. We take a fraction of 20$\%$ as the extinction uncertainty for error propagation purposes. The G band absolute magnitude of the Sun is 4.78, while the bolometric magnitude of the Sun is 4.74 \citep{GaiaDr2}. LAMOST classifies the spectral type of the detected subgiant as G5. The bolometric correction coefficient (BC) is applied as -0.02 according to the empirical relation for Gaia data \citep{bolometric}. 
We use the Vega system for optical magnitude calculations in this work. The mass error is obtained from error propagation. From the relationships and observations above, distance, luminosity and radius are 6.7$^{+1.0}_{-0.8}$kpc, 88$\pm$25L$_{\odot}$ and 15.5$\pm$2.5R$_{\odot}$, respectively.

We compare $M_{1}$ and $L$ with the result derived from stellar evolution models \citep{Bressan2012} and obtain a consistent result \footnote{isochrones tools: \url{http://stev.oapd.inaf.it/cgi-bin/cmd}}. The age is 10$^{8.75}$ years given the stellar parameters obtained. We note that the mass helps us to exclude the possibility of regarding the detected component as a red clump giant which is usually smaller than 2M$_{\odot}$ and rarely larger than 2.5M$_{\odot}$ \citep{Wu2019,Girardi2016}.

The available observations reveal a brighter energy distribution in the redder band which is typical of a subgiant. We find no objects recorded in LTD064402+245919's neighborhood in the SIMBAD database and in X-ray source catalogs, e.g., the Chandra source catalog \citep{Chandra2010}, XMM-Newton Source Catalog \citep{Watson2009}, the ROSAT All-Sky Survey Point Source Catalog \citep{ROSAT2000}. LTD064402+245919 is not catalogued in the LAMOST UV emission catalogs \citep[built from GALEX survey;][]{Bianchi2017, Bai2018}. Infrared observations reveal a brighter magnitude in redder bands, e.g., 12.739 at 2MASS J band, 12.009 at 2MASS K band, 11.883 at WISE W1 band, 8.86 at W4 band \citep{2mass, wise}.

No significant second light is found in the LAMOST-TD spectra (in Fig. \ref{image:spectra}). The second light (if present) would be identified from possible line features caused by different stellar types, distinguishable double lines caused by the Doppler shift, or significant asymmetric line profiles caused by double lines (details in Sec. 4.1). We therefore infer the system is a single-line spectroscopic binary under the distinguishability limitations of LAMOST spectra. We apply a detection limit with the luminosity ratio (undetected component to observed component) as one-third. There would be a very strong tendency to predict a compact object if we reduced the 1/3 detection limit (as described in Sec. 4.1).

\subsection{A combined model using MCMC \label{sec:lq}}

Binary system parameters can be obtained by fitting the light curve and the RV curve using parameterized models \citep{Phoebe2005, Phoebe2020, ELCcodeOrosz}. In this work, we use the PHOEBE routine \citep{Phoebe2020}, developed from the classic WD model \citep{WD}. By considering both the LC and RV curves, the binary system companions are spatially close with well circularized orbits. We, therefore, use a zero eccentricity value in our modeling.

The parameter probability distributions are obtained using the Monte-Carlo Markov Chain (MCMC) method when fitting the model \citep{mcmcbook,emcee}. Parameter interactions make probability distributions not straightforward to quantify. 
The complexity of parameter dependencies leads to the need for a multi-dimension sampling method, e.g., MCMC (emcee) and Multinest (PyMultiNest) routines \citep{emcee, pymultinest}. We obtain our results in this work mainly using the emcee routine and find similar results when applying the PyMultiNest routine.

We fit a combined model to both the radial velocity and the light curves (r band). Additional constraints for the primary star mass, and the mass function are also implemented. A unified likelihood function is generated, taking $\chi^{2}$ as the log-likelihood. The likelihood weights between different constraints are all set equal. The combined model avoids the issue of possibly under-utilizing information or over-emphasizing some of the parameters, compared to using separate, independent models.

Our combined model also includes a detection limit on the undetected component's brightness. We utilize a likelihood cut-off when the brightness ratio of the undetected component to the detected star is greater than one-third.

The free parameters of our combined model are inclination ($i$), central transit time ($T_{0}$), semi-major axis ($a$), mass ratio ($q$), constant offset in radial velocity ($\gamma$), and the RV semi-amplitude ($K$). The priors for these free parameters are all uniform, distributed across the whole possible parameter space. Other parameters are constrained by the PHOEBE model, the radial velocity model, the mass function, the luminosity ratio, and the spectra parallax prediction.

The undetected component is treated as an object of unknown type, e.g., a compact object, a main-sequence star, or a subgiant star. Luminous and dark companions require different setups for PHOEBE modeling. We perform separate combined-model fitting runs for both the Luminous and dark companion cases. In both cases, the best-fit results from MCMC fitting give the same posterior. However, there may be other possible parameter distributions that are less well favored by the data but would be acceptable, e.g., fixing the inclination to different values. In this scenario, luminous and dark companions give different results, mainly due to the non-detection of transits in the light curve (more details are available in Section 3.3).

Any secondary light is less than 1/3 of the total flux, and any influence on the LC shape, e.g., the gravity darkening, limb darkening, or albedo is a significantly smaller-quantity by comparison with the total secondary light. Such modifications from the undetected component do not significantly influence the light curve shape.

Limb darkening is interpolated from PHOEBE's built-in tables. The limb darkening model is commonly following one of the linear, square root, quadratic, or non-linear limb darkening laws \citep{claret2011, YangAtmos, YangLD}. We choose a square-root limb darkening law with parameters supplied by PHOEBE \citep{Phoebe2020}. The gravity darkening exponent and bolometric albedo use convective values of 0.32 and 0.5, respectively, for the observed component. The values are not fixed for the undetected component. Both radiative and convective values are allowed in the fitting.

\begin{figure}[htb]
  \centering
  \includegraphics[width=3.2in]{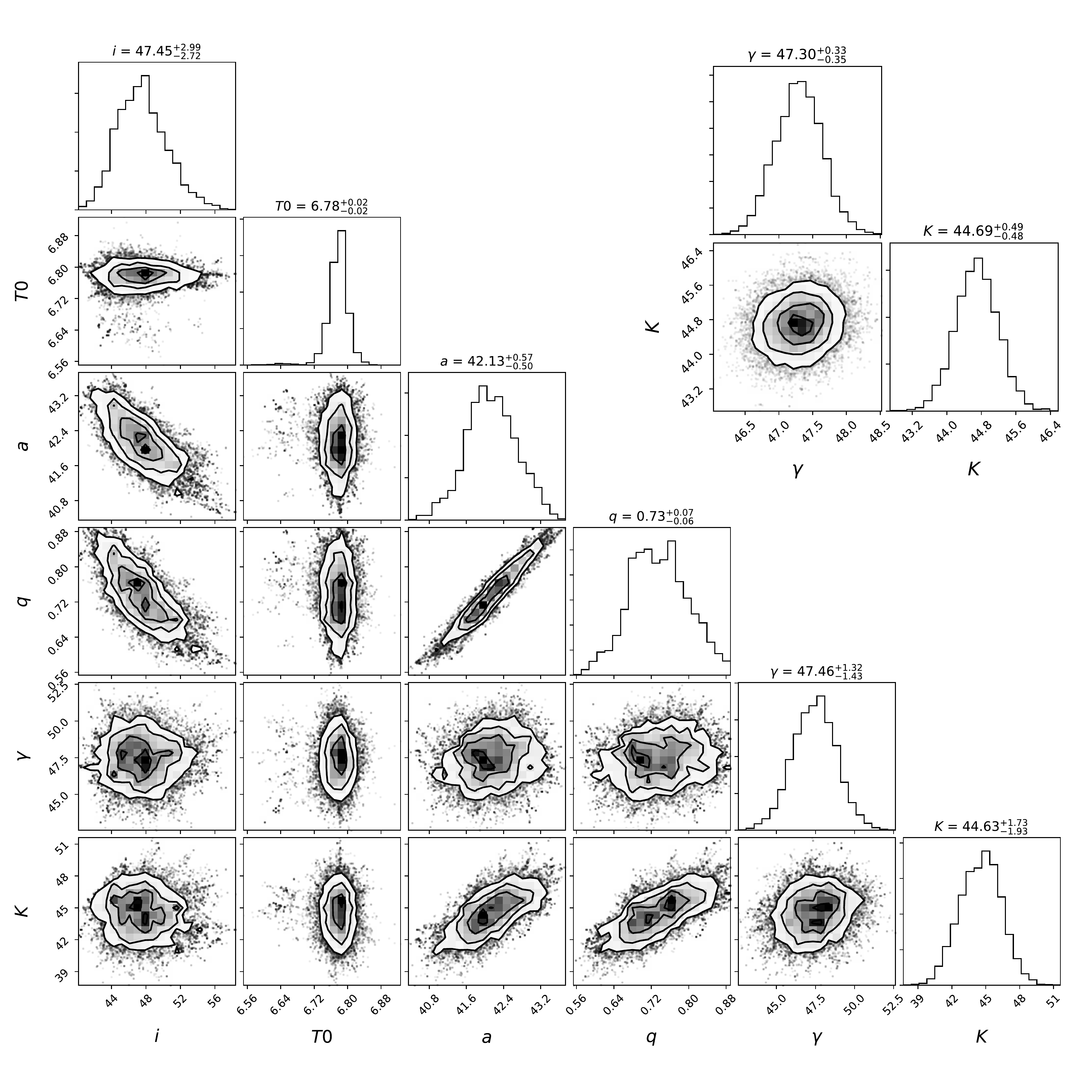}
  \caption{Parameter posterior distributions. The major panel gives results from the combined model. The median values and 1$\sigma$ significance levels are shown in the labels. The upper-right panel shows the $\gamma$ and $K$ posteriors from modeling the radial velocity curve by itself. The $\gamma$ and $K$ distributions are consistent with the results of combined model fit.}
  \label{image:MCMC}
\end{figure}

\begin{figure}[htb]
  \centering
  \includegraphics[width=3.5in]{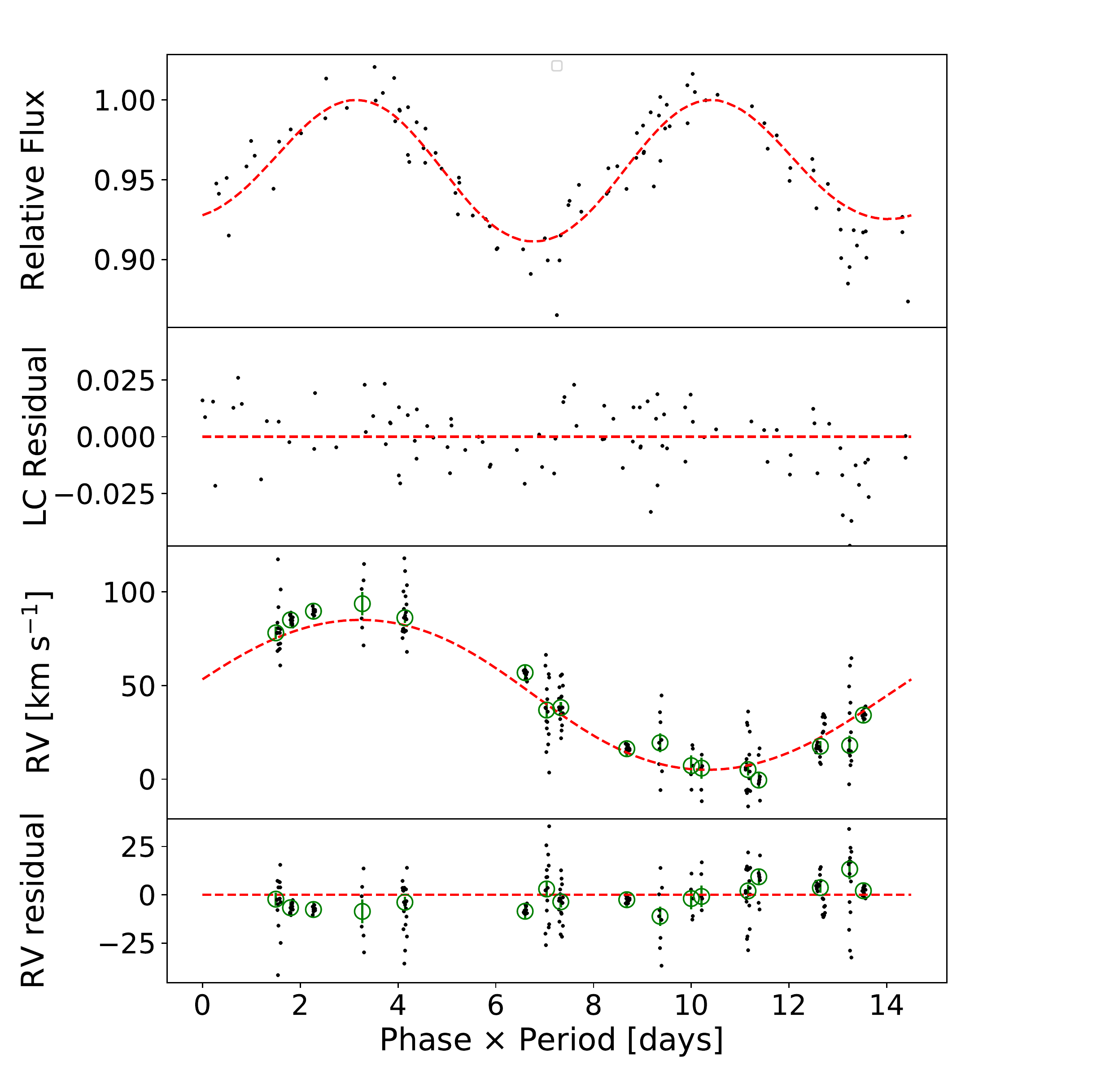}
  \caption{The light curve (top panel), the light curve residuals (second top), the RV curve (third top), and the RV curve residuals (bottom). Black points indicate the observed data. The red lines are the fitted models to the observations (black points). The green circles and error bars indicate binned RV values on the same night. The green circle values are not used in the fitting process. The $\chi^2$ value obtained is 79.}
  \label{image:result}
\end{figure}

\begin{figure}[htb]
  \centering
  \includegraphics[width=3.5in]{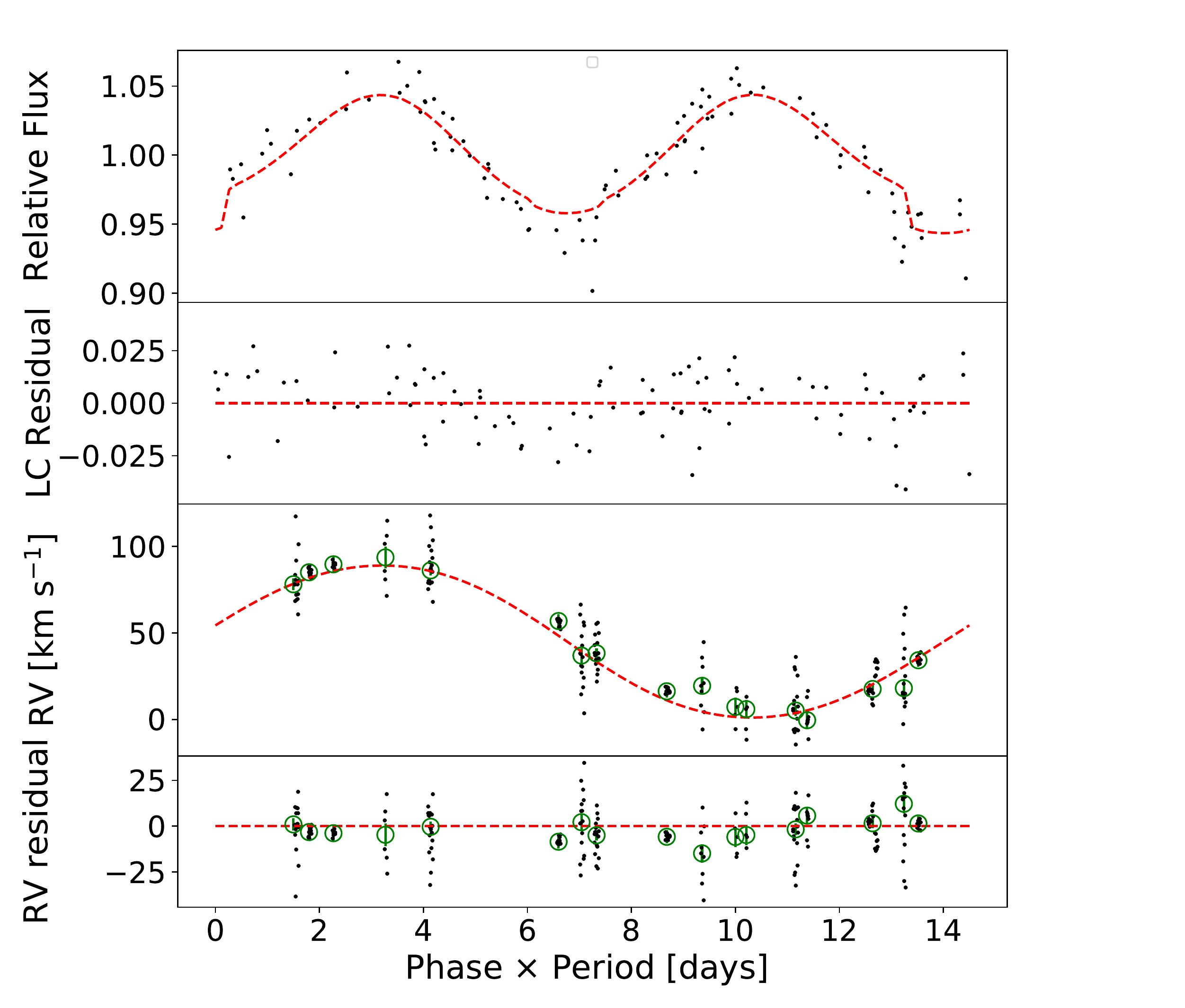}
  \caption{The same plot as Figure \ref{image:result}, but with a fixed inclination of 75 degrees and a main-sequence secondary component. The $\chi^2$ value is 85.}
  \label{image:result75}
\end{figure}

\subsection{Modeling results: Best-fit with Free Inclination}

We run our MCMC modeling process for 80,000 steps, ignoring the first 30,000 steps as burn-in. The results for the monitored free parameters are shown in Fig. \ref{image:MCMC}, which gives the probability distributions and pairwise dependencies of the parameters.

The model light curve, RV curve, and their residuals are shown in Fig. \ref{image:result}. The $\chi^{2}$ value from the fitting process is 79 when we apply a free inclination during fitting.
The residuals of single exposure radial velocities give a standard deviation (SingleStd) of 12 km s$^{-1}$. LTD064402+245919 has an orbital period of 14.50 days so it is reasonable to co-add the RVs on the same night (17 nights in total). The co-added uncertainty is taken as the SingleStd divided by the square root of the number of RVs for the night, according to the rules for error propagation \citep{yangagn}. The median value of the co-added RV uncertainties is 3.9 km s$^{-1}$, while the standard deviation of the co-added RV residuals is 6.3 km s$^{-1}$ (as shown in Fig. \ref{image:result}).

Given the uncertainty of the detected component mass and q, the mass of the secondary component is estimated as 2.02$\pm$0.49M$_{\odot}$. The combined fitting does not distinguish the type of the secondary component. The modeling rules out the possibility of a subgiant as the secondary component. The secondary component would be distinguished with a $q$ of 0.73$^{+0.07}_{-0.06}$ in LAMOST-TD spectra if the secondary component is a subgiant (as discussed in Sec. 4.1).

The detected component stellar surface potential is modeled as 2.38$\pm$0.21dex when filling its Roche lobe. Given that the spectra obtained log\emph{g} is 2.50$\pm$0.25dex, we infer that the detected component has filled its Roche lobe. We note that the filling factor is not set as a fixed parameter. The combined model regards the binary as a detached system and calculates the filling factor.

We compare the posteriors of $K$ and $\gamma$ from our combined model and the model applied to the RV curve only. The posteriors would be significantly different if the observations have any evident inconsistency. The test yields a negligible difference in $K$ and $\gamma$ between the combined model and the RV curve only model, as shown in Fig. \ref{image:MCMC}.

\subsection{Modeling with Fixed Inclinations}

Although our best fit model predicts the parameter distributions, other possible parameter values should be evaluated since the fitted result is highly sensitive to PHOEBE systematics and abnormal data in the light curve. Also, observations, at certain phases, might significantly influence the derived parameter distributions. Particularly important to this work, they might indicate a non `compact object' as the secondary component and a large inclination if transit effects are present. The transit effects occur when the secondary component radius is comparable to the primary component's. A compact object as the secondary component would not cause any transit effects, due to its small radius. In our case, our data analysis does not indicate the presence of transit effects.

There is no significant difference between the two minimum brightness phases in both the r and g bands, indicating that no obvious transit happens (as shown in Figure \ref{image:lc}). We fix the inclination to be, e.g., $\sim$ 75 degrees and refit the data with the combined model. We note that the model light curve is different when the inclination is 75 degrees, depending on whether the undetected companion is a main-sequence star or a compact object. A main-sequence star companion implies a transit should be detected by the LC. The transit is visible at phase 1, inferred from fitting the r band LC (as shown in Figure \ref{image:result75}, first panel). However, the g band light curve yields a deeper decrease at phase 0.5 than phase 1, according to Figure \ref{image:lc}. This inconsistency means there is no clear evidence of any transit occurring in the observed light curve.
In addition, the $\chi^{2}$ value is 85 when fixing the inclination to 75 degrees. This indicates that such an inclination is not well supported by the observations available.

The inclination is a key parameter in the combined model, constrained by both LC and RV observations. The inclination helps determine many other system parameters, most importantly, mass ratio $q$.
The best fit result gives an inclination of 47.45$^{+2.99}_{-2.72}$ degrees with a $\chi^2$ value of 79. Other parameter sets obtained by fixing some parameters are acceptable given the data though they might not be as favored. This is common in transit fitting and leads to the `best-fit' parameter sets \citep[as discussed in][]{YangAtmos, YangLD}. The transit parameters from fixing the inclination can be many $\sigma$ away from the best-fit result. This scenario relates to the model-dependent Bayesian statistics \citep{YangAtmos, YangLD}. In this work, fixing the inclination to, e.g., 75 degrees, supports a model hypothesis that the secondary component is a compact object (due to transit non-detection).

\begin{figure}[htb]
  \centering
  \includegraphics[width=3.5in]{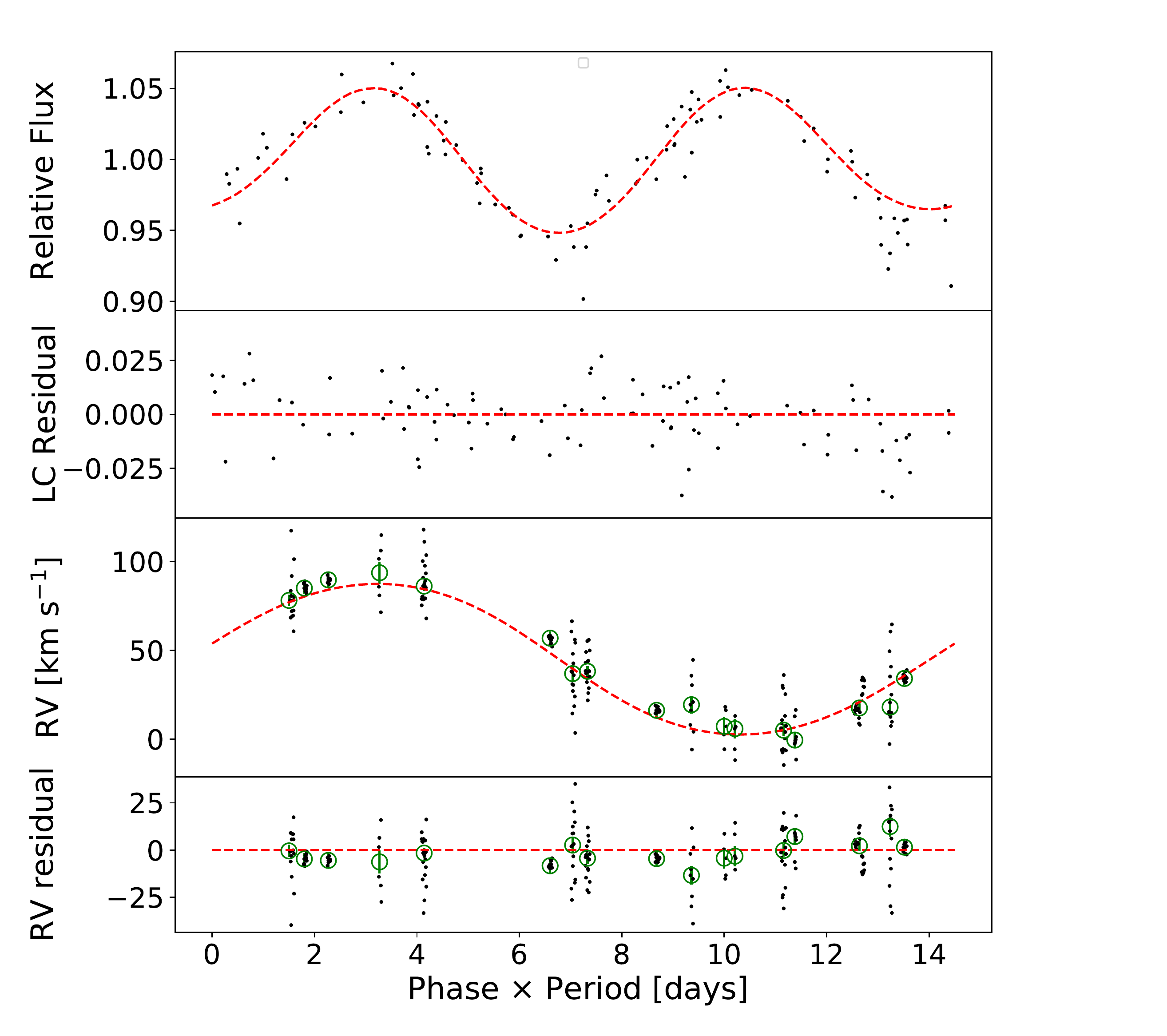}
  \caption{The same plot as Figure \ref{image:result}, but with a fixed inclination of 75 degrees and a zero emission secondary component. The $\chi^2$ value is 85. The model difference appears at transit duration (in the top panel)  when compared to the case in Figure \ref{image:result75}.}
  \label{image:result75compact}
\end{figure}

By fixing the inclination to certain values, we explore other possible parameter sets. We model with an inclination of 75 degrees and a compact object as the secondary companion. The $\chi^2$ value is 85, indicating an acceptable but not a best fit, the same outcome as using a main-sequence companion. The Roche lobe filling factor is 0.88 at this inclination. By comparison, the filling factor is 0.99 when the inclination is 47.45 degrees from the best-fit. We also fix the inclination to 60 degrees with the both main-sequence and compact objects as the secondary companion. The modeling obtains a $\chi^2$ value of 85 and a filling factor of 0.94. The model does not detect any transit with such inclination.

Fixing inclination to 30 degrees is not well supported by the observations, particularly the light curve. A smaller inclination and Roche lobe filling factor lead to a smaller LC ellipsoidal amplitude. However, the amplitude enhancement is not significant when the filling factor is close to 1. LTD064402+245919's primary component nearly fills the Roche lobe, according to the results above. An inclination of 30 degrees requires a larger ellipsoidal amplitude when modeling the light curve which is not possible.

Fixing the inclination to different values has a strong effect on other parameters. The $q$ would be 0.50$^{+0.03}_{-0.04}$ when the inclination is fixed to 75 degrees. This would result in a secondary mass of 1.44$\pm$0.35M$_{\odot}$. Similarly, the secondary mass would be 1.37$\pm$0.34M$_{\odot}$ when the inclination is 90 degrees. This mass is $\sim$ 1$\sigma$ away from the mass obtained from the best-fit. We note that these values are based on a compact object as the secondary component, due to no transit being detected.

We conclude that the fixed inclination results are of interest even though the $\chi^2$ values are larger than best-fit, when the sensitivity of the results to the systematics in the light curve is taken into account. However, with the transit non-detection constraint, the inclination can not be as large as 75 degrees if the secondary component is not set as a compact object. We regard the parameter distribution from the best-fit model as reasonable (giving the companion mass as 2.02$\pm$0.49M$_{\odot}$) when the secondary component type is not set in advance.

In the case of a compact object as the secondary component, working with larger inclinations could be reasonable but the parameter probability space would be significantly enlarged. Solving such an issue is beyond the scope of this work. As a substitute, it is reasonable to give individual parameter distributions depending on different inclinations. The secondary mass is larger than 1M$_{\odot}$ (the 1 $\sigma$ lower limit at 90 degree inclination) when the secondary component is a compact object such as a BH or NS.

\section{Discussion}

We estimate the mass of the undetected component from the combined model using the data available. However, the result does not shed light on the object type of this unseen component. A 2.02$\pm$0.49M$_{\odot}$ component could be an A-type to F-type star, a stripped star evolving from a more massive star, a star enlarging due to mass acquired from the subgiant, or a compact object. Among the possibilities, a stripped star and an enlarging star are overly model-dependent and are difficult to confirm as in other stellar-mass black hole identification works \citep{Shao2020}. Analyzing the possibility of these rare situations is beyond the scope of this work. We narrow the type of the undetected component to be either a main-sequence star (A3V-type), or a compact object.

We present the brightness in different bands in Sec. 3. For LTD064402+245919, the UV band brightness should be observable by GALEX if the secondary companion is a main-sequence star that is hotter than the primary companion (subgiant). We calculate the theoretical GALEX AB magnitude under the assumption of an A3V-type star as the secondary component. The model flux is taken from the ATLAS library \citep{Kurucz1993}. Far-ultraviolet (FUV, 1350–1750\AA) and near-ultraviolet (NUV 1750–2750\AA) GALEX magnitudes should be 19.8 and 18.8, respectively. Following the instructions of \citet{Bianchi2017}, LTD064402+245919 does not match any GALEX observations within 15 arcmins in MAST\footnote{\url{https://mast.stsci.edu/portal/Mashup/Clients/Mast/Portal.html}}. There are GALEX observations when setting the cross-match radius to 30 arcmins. Regarding the Galactic coordinates of 89.67660874 and 9.58877066 degrees, LTD064402+245919 falls into the GALEX observational gaps on the Galactic plane \citep{GALEXoverview, Bianchi2017}.

\subsection{Binary light distinguishing}

A straightforward method for secondary light identification is to shift the spectra to different phases, analyzing the presence of uniformity caused by any binarity \citep[$\lambda$-space method;][]{Ilijic2004, Beck2014, El-Badry2018}. A Fourier transformation based, spectra disentangling method \citep[FDBinary;][]{Ilijic2004} has been put forward\footnote{FDBinary code is available at: \url{http://sail.zpf.fer.hr/cres/}} to identify any binarity. 
The method is applied to multi-epoch, high signal to noise ratio spectra, e.g., SNR $\sim$30 with R$\sim$20000. \citet{Ilijic2004} also discussed the $\lambda$-space method and concluded the two methods are complementary. Using the FDBinary code, \citet{Beck2014} reported disentangling binarity with a detection limit of 3\% of the light ratio. They used 60 observations from the Hermes spectrograph mounted on the 1.2m Mercator Telescope. The resolving power was R = 86,000, with SNR $\sim$20-30.


In this work, we apply the $\lambda$-space method to try and distinguish the secondary light. Line profiles will be different among the phases if there is a secondary light. Line profiles at phases 0.25 and 0.75 present the largest asymmetry if secondary light present, while line profiles at phases 0 and 0.5 reveal the smallest uniformity. 

Only MR spectra are suitable for the secondary light identification. Low resolution spectra do not sample frequently. The blue camera spectra contain no strong lines and have low SNR. We have 37 red camera spectra with a median SNR of 12. The equivalent SNR after shift-adding the spectra is $\sim$ 45. With the spectra available, we do not find any significant binarity signal. This non-detection could be due to the detection limit or to the absence of secondary light. The detection limit is investigated via Monte Carlo simulation.

\begin{figure}[htb]
  \centering
  \includegraphics[width=3.5in,height=3in]{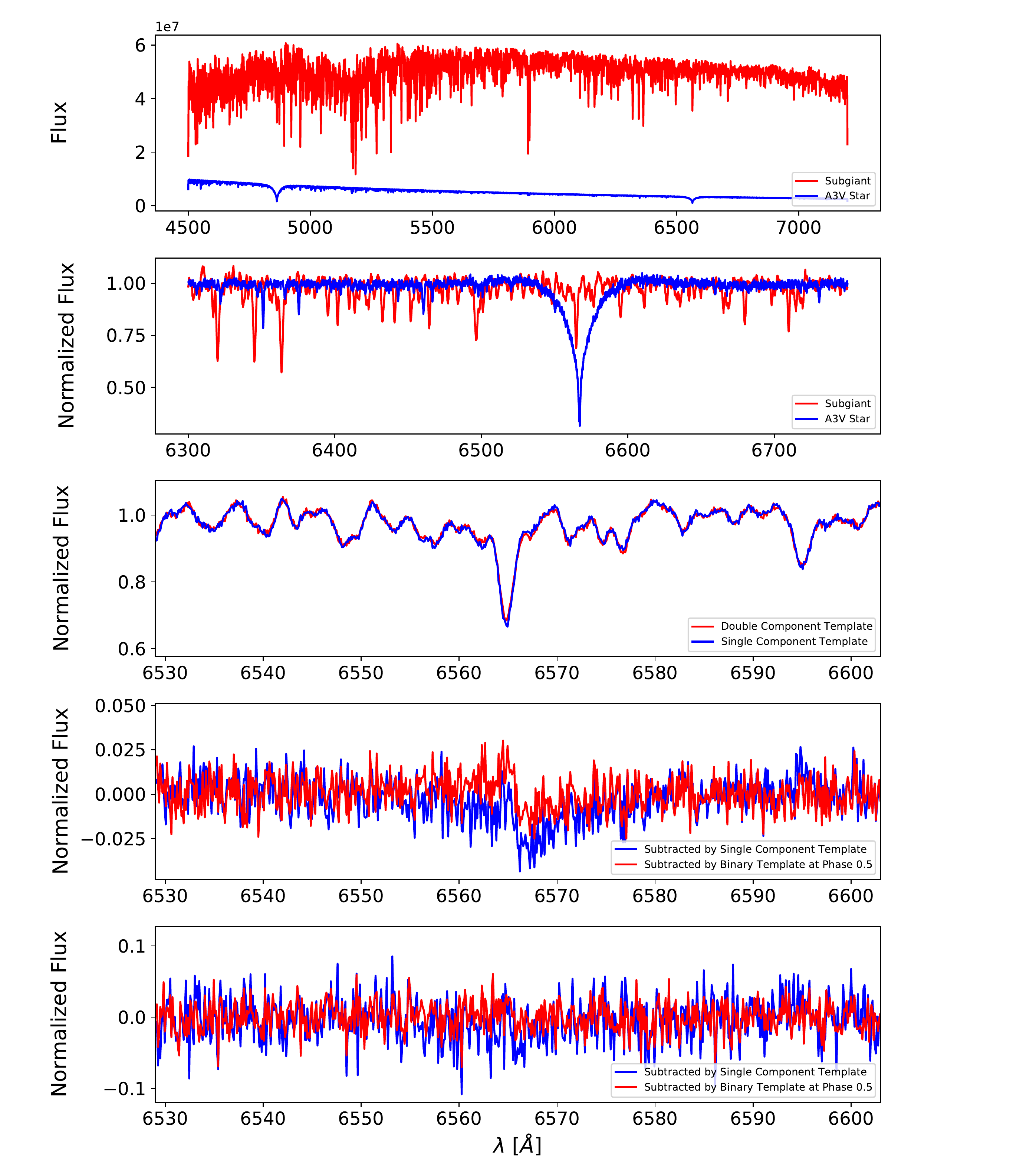}
  \caption{Binarity identification with simulated spectra (R$\sim$7500, SNR $\sim$140). The top panel shows the template spectra of a subgiant and an A3V type star as described in the text. The second panel shows the normalized spectra in which the line structures of the A3V star are clarified. The third panel shows the spectra of double component (red) and single component (blue). The fourth panel gives the double component spectra subtracted by double component spectra at phase 0.5 (red) and single component spectra at phase 0.25 (blue). The bottom panel presents the same subtraction as the fourth panel but with the spectra SNR $\sim$ 45.}
 \label{image:distentangling}
\end{figure}

In the simulation, we first generate template spectra for the system. The basic template is for a subgiant (as shown in Fig. \ref{image:distentangling}). The stellar parameters we use are from the modeling results of the observed component. The undetected component is taken as an A3V-type star. A3V star is the maximum likelihood assessment (for 2.02 M$_{\odot}$) if the undetected object is a main-sequence star. The input stellar parameters to our simulation are $T_{\rm eff}$=8750K, log $g$=4.27dex, [Fe/H]=-0.54dex, vsini=2.34 km s$^{-1}$, mass=2.02M$_{\odot}$, and radius=1.7R$_{\odot}$.

The stellar spectra template is from \citet{Prieto2018}, sampled at the LAMOST median resolution (R=7500) with an SNR of 140. The SNR is achieved when co-adding 10 times more spectra in this work. With this SNR, the binarity, if present, would be distinguishable at 5$\sigma$ level. The luminosity ratio of the sources is 1:9.45. The line features of the A3V-type star are severely obscured when combined with the subgiant spectra. The H$\alpha$ line of the A3V-type star is much stronger than other lines, among LAMOST spectra. For the subgiant, the H$\alpha$ line is not significantly stronger than other lines. The spectral difference due to binarity (if present) should thus be dominated by H$\alpha$. Co-adding weak lines does not significantly improve the distinguishing evidence but acts to smooth the H$\alpha$ feature.

Using the strongly featured H$\alpha$ line for the binarity detection, the two components have a line center difference of 107 km s$^{-1}$ (as shown in the second panel, Fig. \ref{image:distentangling}). The spectra at phase 0.25 and 0.75 reveal the largest difference between two components and one component, and would show sinusoidal subtraction residuals if secondary light is present (as shown in the third and fourth top panels, Fig. \ref{image:distentangling}). The subtraction between observation and the single component template is strongly model-dependent. The comparison however between the spectra at phase 0.25 (0.75) and phase 0 (0.5) obtains a similar structure but is more model-independent (as shown in the fourth top panel, Fig.\ref{image:distentangling}). Any significant difference is defined by
the difference between the median value of bumps on each side of the sinusoidal curve.

Binarity is not detectable with the spectra available, as shown in the bottom panel of Fig. \ref{image:distentangling}. The sinusoidal structure is detectable when the SNR reaches $\sim$140 through shifting and co-adding the spectra. It is also detectable with spectra available if we increase the secondary brightness to be 1/3 of the primary light. The detection significance when the brightness ratio is 1/3 or when having 10 times more spectra, 
is similar, in agreement with the simulation result.

The detection ability in our work is similar to the result in the work of \citet{Beck2014}. The equivalent spectra quality of R$\sim$7500, SNR $\sim$140 under the conditions of \citet{Beck2014} needs 4 observations to acquire. There are 60 observational epochs in \citet{Beck2014} which means that the detection limit of luminosity ratio should reach 3.8\%. For comparison, the reported detection limit in \citet{Beck2014} is 3\%. Considering the observation phase coverage and spectra quality variances, the detection limits in our work and \citet{Beck2014} are consistent.

The detection limit reported by \citet{El-Badry2018} enabled a main-sequence secondary component with a period of $\sim$ 10 days and a mass ratio of $\sim$ 0.05 to be detected. According to classic mass-luminosity relations, a secondary component with a luminosity ratio of 3$\times$$10^{-5}$ would be detectable by \citet{El-Badry2018}. Regarding the high SNR APOGEE data (SNR=200, R=22,500) used by \citet{El-Badry2018}, the detection ability of \citet{El-Badry2018} exceeds more than 200 times the similar (equivalent) data conditions of \citet{Beck2014}. It means that \citet{Beck2014} could detect signals at a significance level of 3$\times$$10^{-5}$$\times$$N_{lines}$$\times$SNR, where $N_{lines}$ is the number of spectral lines used in the code to distinguish components. A detection significance level of 1$\sigma$ needs 1500 spectra lines to be achieved. Further detection limit discussions are beyond the scope of this work.


\section{Conclusion}

We have identified a binary system LTD064402+245919, with a detected subgiant star as the primary component (2.77$\pm$0.68M$_{\odot}$), with a full Roche lobe. The secondary component is undetected in brightness but has an estimated mass of 2.02$\pm$0.49M$_{\odot}$ assuming the companion is not a compact object. The mass of the secondary companion is larger than 1M$_{\odot}$ under a compact object assumption. The difference is attributed to transit non-detection in the light curves. 
X-ray surveys covering the sky area do not identify the system as an X-ray source. The observational data suggest a binary system with an unknown secondary companion. The secondary component could be an A-type star or a compact object quiescent in X-ray.

Identification of LTD064402+245919 makes use of LAMOST-TD and ZTF. We have built a machine learning based pipeline to identify the photometric variable stars in the LAMOST-TD sources. The pipeline identifies LTD064402+245919 as a contact binary with an orbital period of 14.50 days. 107 spectra from LAMOST-TD are used for the RV curve fitting and 102 photometric measurements in the r band are used for LC fitting.

The RV and light curves are fitted using a combined MCMC model that includes an RV curve model, an LC model, a primary stellar model, a binary mass function, and a detection limit on the secondary light. The key parameters from the modeling are orbital inclination, mass ratio, RV semi-amplitude with values 47.45$^{+2.99}_{-2.72}$ degrees, 0.73$^{+0.07}_{-0.06}$, and 44.63$^{+1.73}_{-1.93}$ km s$^{-1}$, respectively, when the secondary component type is not set. We do not give global estimates of the parameter distributions when setting the secondary companion as a compact object. Fixing inclination to specific values can be regarded as adding model hypotheses. Combining models with different hypotheses into a global estimation is a common difficulty in Bayesian statistics \citep{YangAtmos,YangLD}.

The detection limit for the secondary light is crucial for identifying the type of undetected source. We have examined the ability to distinguish secondary light by using Monte Carlo simulation. The simulation indicates a secondary component should be detectable if the luminosity ratio is higher than 1:3. The simulation also shows that a secondary component (if present) should be detectable if the number of spectra is increased tenfold (equivalent to the SNR increasing threefold).

Our work provides a possible candidate system for compact object searches. The identification algorithm and software could be used as general-purpose tools when looking for X-ray quiescent compact objects by analyzing single-line spectroscopic binaries. Utilizing photometric light curves, our method is especially effective when spectroscopic observations are not sufficient to monitor the complete binary system rotation process.

\section*{acknowledgements}
We highly appreciate the constructive suggestions from the referee which greatly improve the quality of the paper. The authors thank Wei-Ming Gu and Hao-Tong Zhang for their instructive feedback. Fan Yang would like to thank Chao Liu and Chang-Qing Luo for helpful discussions, and WeiKai Zong for advice on photometric data reduction. 
Bo Zhang thanks the LAMOST FELLOWSHIP funding. The LAMOST FELLOWSHIP is supported by Special Funding for Advanced Users, budgeted and administrated by Center for Astronomical Mega-Science, Chinese Academy of Sciences (CAMS).
Fan Yang, Bo Zhang, Su-Su Shan, and Ji-Feng Liu acknowledge funding from the Cultivation Project for LAMOST Scientific Payoff and Research Achievement of CAMS-CAS, the National Key Research and Development Program of China (No.2016YFA0400800), and the National Natural Science Foundation of China (NSFC) (No.11988101). Xing Wei is supported by the NSFC (No. 11872246, 12041301) and the Beijing Natural Science Foundation (No. 1202015). JNF acknowledges support from the NSFC through grants 11833002, 12090040 and 12090042.

Guoshoujing Telescope (the Large Sky Area Multi-Object Fiber Spectroscopic Telescope LAMOST) is a National Major Scientific Project built by the Chinese Academy of Sciences. Funding for the project has been provided by the National Development and Reform Commission. LAMOST is operated and managed by the National Astronomical Observatories, Chinese Academy of Sciences. Based on observations obtained with the Samuel Oschin Telescope 48-inch and the 60-inch Telescope at the Palomar Observatory as part of the Zwicky Transient Facility project. Major funding has been provided by the U.S National Science Foundation under Grant No. AST-1440341 and by the ZTF partner institutions.

\bibliographystyle{mnras}
\bibliography{ref}
\end{document}